\documentclass[article,12pt,superscriptaddress,showkeys,showpacs] {revtex4}
\usepackage{amsfonts,bm}
\usepackage{graphicx, amsmath}
\usepackage{multirow}
\begin{document}
\title[Short Title]{Multi-qubit non-adiabatic holonomic controlled quantum gates in decoherence-free subspaces}
\author{Shi Hu}
\affiliation{Department of Physics, College of Science, Yanbian
University, Yanji, Jilin 133002, People's Republic of China}
\author{Wen-Xue Cui}
\affiliation{Department of Physics, College of Science, Yanbian
University, Yanji, Jilin 133002, People's Republic of China}
\author{Qi Guo}
\affiliation{College of Physics and Electronics Engineering, Shanxi University, Taiyuan 030006, People's Republic of China}
\author{Hong-Fu Wang\footnote{E-mail: hfwang@ybu.edu.cn}}
\affiliation{Department of Physics, College of Science, Yanbian
University, Yanji, Jilin 133002, People's Republic of China}
\author{Ai-Dong Zhu}
\affiliation{Department of Physics, College of Science, Yanbian
University, Yanji, Jilin 133002, People's Republic of China}
\author{Shou Zhang\footnote{E-mail: szhang@ybu.edu.cn}}
\affiliation{Department of Physics, College of Science, Yanbian
University, Yanji, Jilin 133002, People's Republic of China}

\begin{abstract}
Non-adiabatic holonomic quantum gate in decoherence-free subspaces is of greatly practical importance due to its built-in fault tolerance, coherence stabilization virtues, and short run-time. Here we propose some compact schemes to implement two- and three-qubit controlled unitary quantum gates and Fredkin gate. For the controlled unitary quantum gates, the unitary operator acting on the target qubit is an arbitrary single-qubit gate operation. The controlled quantum gates can be directly implemented using non-adiabatic holonomy in decoherence-free subspaces and the required resource for the decoherence-free subspace encoding is minimal by using only two neighboring physical qubits undergoing collective dephasing to encode a logical qubit.
\end{abstract}

\maketitle \section{Introduction}\label{sec0}
Based on the quantum parallelism, quantum computation is believed to can speed up the solution of a number of mathematical tasks and has attracted more and more interests. The key step to implement effective quantum computation is the construction of robust quantum gates. Holonomic quantum computation (HQC), which is first proposed by Zanardi and Rasetti~\cite{PMPLA9994} basing on adiabatic evolution, is regarded as a promising way to implement universal sets of robust gates. It can be robust against certain types of errors in the control process and has been used to realize robust quantum computation~\cite{JVAGN00403,LJP033,XM0187,SZPRL0391,LPDPRL0595,LZSPRA0674,
XQZPRA0674,XCHCPRL09103,VMDPRA1081} by taking advantage of non-Abelian geometric phases~\cite{FAPRL8452} which only depend on global geometric properties of the evolution paths.
Unfortunately, however, the long run-time requirement for the desired parametric control associated with adiabatic evolution makes the quantum gates become vulnerable to open system effects and parameter fluctuations that may lead to loss of coherence. In order to remove the problem of long run-time associated with the original form of HQC~\cite{PMPLA9994}, Sj\"{o}qvist {\it et al.} developed a non-adiabatic generalization of HQC~\cite{EDLBMKNJP1214} in which high-speed universal quantum gates can be implemented using non-adiabatic non-Abelian geometric phases~\cite{JPLA8813}. Non-adiabatic HQC has also been experimentally demonstrated in different physical systems, such as three-level transmon qubit~\cite{AJKMSASN13496}, nuclear magnetic resonance (NMR) quantum information processor~\cite{GGGPRL13110}, and diamond nitrogen-vacancy centers~\cite{SASGNC145,CWLWCFLN14514}.

Besides errors from the control of quantum system, decoherence, arised from the inevitable interaction between the quantum system and environment, is another main challenge in implementing robust quantum gates. Decoherence will destruct the desired coherence of the system, so it is harmful for effective quantum computation. One of the promising strategies to avoid decoherence is decoherence-free subspaces (DFSs) which utilize the symmetry structure of the system-environment interaction~\cite{DIK9881}. The basic idea of DFS is that information encoded in it still  undergoes unitary evolution even though taking the decoherence caused by environment into account. In addition, DFSs have been experimentally demonstrated in a host of physical systems~\cite{PAJAS00290,DVMCWCDS01291,MJKAPRL0391,JDLPRL0391,
MMSCAHPRL0492}.

Many efforts have been devoted to combining the fault tolerance of HQC and the quantum coherence stabilization virtues of DFSs~\cite{LPDPRL0595,LZSPRA0674,XQZPRA0674}. In 2005, Wu {\it et al.}~\cite{LPDPRL0595} implemented HQC in DFSs which was robust against some stochastic errors and collective dephasing. However, the long run-time associated with the adiabatical control of the parameters and the using of four neighboring physical qubits undergoing collective dephasing to encode a logical qubit are big challenges in experiment. After that, Xu {\it et al.}~\cite{GJDELPRL12109} developed a non-adiabatic generalization of HQC in DFSs which could overcome the long run-time requirement of its adiabatic counterpart. Latter, some other schemes for non-adiabatic HQC in DFSs in different physical systems have also been proposed~\cite{ZYWZHPRA1489,JWYZOE1523,ZJZPRA1592}. However, all the above schemes only focused on one- and two-qubit gates. As we all known, it is too complex to implement most algorithms with the increase of the number of qubits if only one- and two-qubit gates are available. The direct implementation of multiqubit gates, which is generally believed to provide a simpler design, a faster operation, and a lower decoherence, is thus of greatly practical importance.

In this paper, inspired by above works, we propose some compact schemes to implement non-adiabatic holonomic two- and three-qubit controlled unitary quantum gates and Fredkin gate in DFSs. Here the unitary operator acting on the target qubit in controlled unitary quantum gates, is an arbitrary single-qubit gate operation by varying the parameters independently. These controlled quantum gates can be directly implemented, which avoids the extra work of combining two gates into one. Furthermore, they are robust against certain types of errors in the control process and the decoherence caused by environment, and can be implemented in a high speed. This is the first scheme for implementing three-qubit controlled quantum gates using non-adiabatic holonomy in DFSs. Moreover, an attractive feature of our schemes is that the resources cost for the DFSs encoding is minimal by using only two neighboring physical qubits to encode a logical qubit.

\section{QUANTUM HOLONOMY AND PHYSICAL MODEL}\label{sec1}
We now briefly show how quantum holonomy can arise in non-adiabatic unitary evolution before introducing our physical model. Consider a quantum system described by an $N$-dimensional state space and governed by Hamiltonian $H(t)$. Assume that there is a time-dependent $M$-dimensional subspace $S(t)$ spanned by the orthonormal basis vectors $\{|\psi_{m}(t)\rangle\}_{m=1}^{M}$. The evolution operator $\mathcal{U}(\tau,0)$ is a holonomic matrix acting on $S(0)$ spanned by $\{|\psi_{m}(0)\rangle\}_{m=1}^{M}$ if $|\psi_{m}(t)\rangle$ satisfies the following conditions~\cite{GJDELPRL12109}:
\begin{eqnarray}\label{e01}
(\mathrm{i})\sum_{m=1}^{M}|\psi_{m}(\tau)\rangle\langle\psi_{m}(\tau)|
=\sum_{m=1}^{M}|\psi_{m}(0)\rangle\langle\psi_{m}(0)|,
\end{eqnarray}
\begin{eqnarray}\label{e02}
~(\mathrm{ii})~\langle\psi_{m}(t)|H(t)|\psi_{l}(t)\rangle=0,~~m,l=1,2,...,M,
\end{eqnarray}
where $\tau$ is the evolution period, $|\psi_{m}(t)\rangle=\mathcal{U}(t,0)|\psi_{m}(0)\rangle=
\textbf{T}\mathrm{exp}(-i\int_{0}^{t}H(t')dt')|\psi_{m}(0)\rangle$, \textbf{T} is time ordering. Here condition (i) ensures that the evolution of subspace $S(0)$ is cyclic, while  condition (ii) means that the evolution is purely geometric.

In order to combine the fault tolerance of HQC and the quantum coherence stabilization virtues of DFSs, we consider the following physical model. The quantum system  consists of $N$ physical qubits interacting collectively with a dephasing environment. The interaction between the quantum system and its environment is described by the interaction Hamiltonian
\begin{eqnarray}\label{e03}
H_{I}=\Big(\sum_{k=1}^{N}Z_{k}\Big)\otimes B,
\end{eqnarray}
where $Z_{k}$ is the Pauli $Z$ operator for the $k$th physical qubit and $B$ is an arbitrary environment operator. Due to the symmetry of the interaction we can find a DFS to protect quantum information against decoherence. For the simplest case, i.e., the number of physical qubits is two, there exists a DFS:
\begin{eqnarray}\label{e04}
S^{D}=\mathrm{Span}\{|01\rangle,|10\rangle\}.
\end{eqnarray}
We can use this subspace to encode a logical qubit, i.e., $|0\rangle_{L}=|01\rangle$, $|1\rangle_{L}=|10\rangle$, hereafter we use the subscript $L$ to denote logical states. Obviously, the resources cost for the DFS encoding is minimal by using only two neighboring physical qubits, which undergo collective dephasing to encode a logical qubit. In the following, we will use this encoding to implement controlled quantum gates.

\section{TWO-QUBIT CONTROLLED UNITARY GATE}\label{sec2}
In this section we demonstrate how to implement a non-adiabatic holonomic two-qubit controlled unitary gate, denoted as $C_{1}$-$U$ gate, in DFS. Here $U$ is an arbitrary single-qubit unitary gate operation acting on the target qubit, whose matrix form is given by
\begin{eqnarray}\label{e05}
U=\begin{pmatrix}u_{00}&u_{01}\\
u_{10}&u_{11}
\end{pmatrix}.
\end{eqnarray}
To this end, we consider four physical qubits interacting collectively with the dephasing environment and there exists a six-dimensional DFS:
\begin{eqnarray}\label{e06}
S^{D_{1}}=\mathrm{Span}\Big\{|0101\rangle,|0110\rangle,|1001\rangle,|1010\rangle,
|0011\rangle\,|1100\rangle\Big\}.
\end{eqnarray}
We encode logical qubits in the subspace
\begin{eqnarray}\label{e07}
S^{L_{1}}=\mathrm{Span}\Big\{|0101\rangle,|0110\rangle,|1001\rangle,|1010\rangle\Big\},
\end{eqnarray}
where the logical qubit states are denoted as $|0\rangle_{L}|0\rangle_{L}=|0101\rangle$, $|0\rangle_{L}|1\rangle_{L}=|0110\rangle$,
$|1\rangle_{L}|0\rangle_{L}=|1001\rangle$, and $|1\rangle_{L}|1\rangle_{L}=|1010\rangle$. $S^{L_{1}}$ is a subspace of $S^{D_{1}}$ and the remaining vectors $|0011\rangle$ and $|1100\rangle$ are used as ancillary states, denoted as $|a_{1}\rangle=|0011\rangle$ and $|a_{2}\rangle=|1100\rangle$ for convenience. Under the basis $\{|0\rangle_{L}|0\rangle_{L}$, $|0\rangle_{L}|1\rangle_{L}$,
$|1\rangle_{L}|0\rangle_{L}$, $|1\rangle_{L}|1\rangle_{L}\}$, the $C_{1}$-$U$ gate is written as~\cite{ACRDNPTJHPRA9552}
\begin{eqnarray}\label{e08}
C_1-U=
\begin{pmatrix}~1~~&&0~~&&0~~&0~\\
~0~~&&1~~&&0~~&0~\\
~0~~&&0~~&&u_{00}~~&u_{01}~\\
~0~~&&0~~&&u_{10}~~&u_{11}~\\
\end{pmatrix}.
\end{eqnarray}
In order to implement $C_{1}$-$U$ gate, we consider the following Hamiltonian
\begin{eqnarray}\label{e09}
H_{1}&=&\frac{1}{2}\bigg\{(I_{2}+Z_{2})
\Big[\Delta_{1}(I_{1}+Z_{1})
+(\Omega_{1}R^{x}_{13}
+\Omega_{2}R^{x}_{14}+\mathrm{H.c.})\Big]\cr\cr&&+(I_{1}-Z_{1})
\Big[\Delta_{2}(I_{2}-Z_{2})
+(\Omega_{3}R^{x}_{23}
+\Omega_{4}R^{x}_{24}+\mathrm{H.c.})\Big]\bigg\},
\end{eqnarray}
where $R^{x}_{lm}=\dfrac{1}{4}(X_{l}-iY_{l})(X_{m}+iY_{m})$, $I$ is the one-qubit identity matrix, $X$, $Y$, and $Z$ are Pauli matrices acting on corresponding physical qubit, H.c. means Hermitian conjugate, and $\Delta_{i}$ and $\Omega_{i}$ are controllable coupling parameters, with
\begin{eqnarray}\label{e10}
\Delta_{1}&=&-\Omega\sin\xi,
~~~~~~~~~~~~~~~~~\Delta_{2}~=~-\Omega\sin\gamma,\cr\cr
\Omega_{1}&=&\Omega\cos\xi\cos\frac{\alpha}{2},
~~~~~~~~~~~~\Omega_{3}~=~-\Omega\cos\gamma\cos\frac{\alpha}{2},\cr\cr
\Omega_{2}&=&\Omega e^{i\beta}\cos\xi\sin\frac{\alpha}{2},
~~~~~~~~~\Omega_{4}~=~\Omega e^{i\beta}\cos\gamma\sin\frac{\alpha}{2}.
\end{eqnarray}
The Hamiltonian $H_{1}$ can be rewritten as
\begin{eqnarray}\label{e11}
H_{1}'&=&-2\Omega \big(\sin\xi|a_{1}\rangle\langle a_{1}|
+\sin\gamma|a_{2}\rangle\langle a_{2}|\big)\cr\cr
&&+\Omega\big(\cos\xi|1\rangle_{L}|+\rangle_{L}\langle a_{1}|
+\cos\gamma|1\rangle_{L}|-\rangle_{L}\langle a_{2}|+\mathrm{H.c.}\big),
\end{eqnarray}
where we have used two orthogonal states $|+\rangle_{L}=\cos\dfrac{\alpha}{2}|0\rangle_{L}
+e^{i\beta}\sin\dfrac{\alpha}{2}|1\rangle_{L}$ and $|-\rangle_{L}=e^{-i\beta}\sin\dfrac{\alpha}{2}|0\rangle_{L}
-\cos\dfrac{\alpha}{2}|1\rangle_{L}$. The subspace spanned by $\{|+\rangle_{L}$, $|-\rangle_{L}\}$ is the same as that by $\{|0\rangle_{L}$, $|1\rangle_{L}\}$. The evolution operator associated with $H_{1}$ is $\mathcal{U}_{1}(t)=e^{-iH_{1}t}$. With the choice of $\Omega\tau_{1}=\pi$, the resulting evolution operator is given by
\begin{eqnarray}\label{e12}
\mathcal{U}_{1}(\tau_{1})=
\left(\begin{array}{cccccc}
e^{i(\delta-\frac{\theta}{2})}&0&~~~0~~~&~~~0~~~&0&0\\
0&e^{i(\delta+\frac{\theta}{2})}&0&0&0&0\\
0&0&1&0&0&0\\
0&0&0&1&0&0\\
0&0&0&0&e^{i(\delta-\frac{\theta}{2})}&0\\
0&0&0&0&0&e^{i(\delta+\frac{\theta}{2})}
\end{array}\right),
\end{eqnarray}
in the basis $\{|a_{1}\rangle,$ $|a_{2}\rangle,$ $|0\rangle_{L}|+\rangle_{L},$ $|0\rangle_{L}|-\rangle_{L},$
$|1\rangle_{L}|+\rangle_{L},$ $|1\rangle_{L}|-\rangle_{L}\}$, where $\delta-\theta/2=\pi+\pi\sin\xi$ and
$\delta+\theta/2=\pi+\pi\sin\gamma$. Since the parameters $\xi$ and $\gamma$ are mutually independent, we can vary the parameters $\delta$ and $\theta$ independently.

Therefore, for the states in the logical subspace $S^{L_{1}}$, the action of the evolution operator $\mathcal{U}_{1}(\tau_{1})$ is equivalent to $C_{1}$-$U$ gate and the single-qubit unitary gate operation $U$ is written as
\begin{eqnarray}\label{e13}
U=e^{i(\delta-\frac{\theta}{2})|+\rangle_{L}\langle +|+i(\delta+\frac{\theta}{2})|-\rangle_{L}\langle -|}.
\end{eqnarray}
Under the basis $\{|0\rangle_{L}, |1\rangle_{L}\}$, defining the Pauli operators as $\sigma_{x}=|0\rangle_{L}\langle1|+|1\rangle_{L}\langle0| $, $\sigma_{y}=-i|0\rangle_{L}\langle1|+i|1\rangle_{L}\langle0| $, and
$\sigma_{z}=|0\rangle_{L}\langle0|-|1\rangle_{L}\langle1|$, then $U$ can be rewritten as
\begin{eqnarray}\label{e14}
U=\mathrm{exp}(i\delta)R_{\hat{n}}(\theta),~~~~
R_{\hat{n}}(\theta)=\mathrm{exp}\left(-i\frac{\theta}{2}\hat{n}\cdot\bm{\sigma}\right),
\end{eqnarray}
with $\bm{\sigma}=(\sigma_{x},\sigma_{y},\sigma_{z})$ and the unit vector $\hat{n}=(\sin\alpha\cos\beta, \sin\alpha\sin\beta, \cos\alpha)$. In the above, $R_{\hat{n}}(\theta)$ represents a single-qubit rotation around the direction $\hat{n}$ with angle $\theta$. Thus $U$ corresponds to an arbitrary single-qubit gate operation by varying the parameters $\delta$, $\theta$, $\alpha$, and $\beta$ independently~\cite{MIQCQI}. In particular, when setting $\delta=\theta/2=\alpha=\pi/2$
($\xi=\pi$, $\gamma=0$, $\alpha=\pi/2$) and $\beta=0$, we can implement a two-qubit controlled-NOT (CNOT) gate.

Since $S^{D_{1}}$ is an invariant subspace of the evolution operator, $\mathcal{U}_{1}(\tau_{1})$ has decoherence-free property. Next, we use conditions $(\mathrm{i})$ and $(\mathrm{ii})$ to check that $\mathcal{U}_{1}(\tau_{1})$ is a holonomic matrix acting on $S^{L_{1}}$. For condition $(\mathrm{i})$, the subspace spanned by $\{\mathcal{U}_{1}(\tau_{1})|0\rangle_{L}|0\rangle_{L},$ $\mathcal{U}_{1}(\tau_{1})|0\rangle_{L}|1\rangle_{L},$
$\mathcal{U}_{1}(\tau_{1})|1\rangle_{L}|0\rangle_{L},$ $\mathcal{U}_{1}(\tau_{1})|1\rangle_{L}|1\rangle_{L}\}$ coincides with $S^{L_{1}}$, it is satisfied. While for condition $(\mathrm{ii})$, considering that $\mathcal{U}_{1}(t)$ commutes with $H_{1}$, condition $(\mathrm{ii})$ reduces to $\langle k|H_{1}|k^{'}\rangle=0$, where $|k\rangle, |k^{'}\rangle\in \{|0\rangle_{L}|0\rangle_{L}, |0\rangle_{L}|1\rangle_{L},
|1\rangle_{L}|0\rangle_{L}, |1\rangle_{L}|1\rangle_{L}\}$. From Eq.~({\ref{e11}}), it is easy to find that condition $(\mathrm{ii})$ is also satisfied. Therefore, $\mathcal{U}_{1}(\tau_{1})$ is a holonomic matrix acting on $S^{L_{1}}$ with decoherence-free property.

Through the above illustration, a non-adiabatic holonomic $C_{1}$-$U$ gate in which $U$ is an arbitrary single-qubit gate operation in DFS with two- and three-body interactions have been directly and successfully implemented. It is worth pointing out that one needs four-body interaction~\cite{GJDELPRL12109} or the combination of a single-qubit gate and a two-qubit nontrivial gate~\cite{JWYZOE1523} to implement a non-adiabatic holonomic CNOT gate in DFS.

\section{THREE-QUBIT CONTROLLED unitary GATE}\label{sec1}
It is well known that by using two CNOT gates, two $C_{1}$-$V$ gates $(V^{2}=U)$, and a $C_{1}$-$V^{\dag}$ gate, one can get a three-qubit controlled unitary gate with two control qubits and a unitary operator $U$ acting on a target qubit, which is denoted as $C_{2}$-$U$ gate~\cite{ACRDNPTJHPRA9552}. Obviously, this combination is very complex and it is more desirable to implement $C_{2}$-$U$ gate directly. In this section we will show how to implement the $C_{2}$-$U$ gate directly in DFS. To this end, we need six physical qubits interacting collectively with the dephasing environment to construct a ten-dimensional DFS:
\begin{eqnarray}\label{e15}
S^{D_{2}}=\mathrm{Span}\Big\{|010101\rangle,|010110\rangle,
|011001\rangle,|011010\rangle,|100101\rangle,~\cr\cr
|100110\rangle,|101001\rangle,|101010\rangle,
|100011\rangle,|101100\rangle\Big\}.
\end{eqnarray}
Similar to the case of $C_{1}$-{\it U} gate, we encode logical qubits in the subspace
\begin{eqnarray}\label{e16}
S^{L_{2}}=\mathrm{Span}\Big\{|010101\rangle,|010110\rangle,
|011001\rangle,|011010\rangle,~\cr\cr
|100101\rangle,|100110\rangle,|101001\rangle,|101010\rangle\Big\},
\end{eqnarray}
and the logical qubit states are denoted as
\begin{align}\label{e17}\notag
|0\rangle_{L}|0\rangle_{L}|0\rangle_{L}&=|010101\rangle, &|0\rangle_{L}|0\rangle_{L}|1\rangle_{L}&=|010110\rangle,
\end{align}
\begin{align}\notag
|0\rangle_{L}|1\rangle_{L}|0\rangle_{L}&=|011001\rangle, &|0\rangle_{L}|1\rangle_{L}|1\rangle_{L}&=|011010\rangle,
\end{align}
\begin{align}\notag
|1\rangle_{L}|0\rangle_{L}|0\rangle_{L}&=|100101\rangle, &|1\rangle_{L}|0\rangle_{L}|1\rangle_{L}&=|100110\rangle,
\end{align}
\begin{align}
|1\rangle_{L}|1\rangle_{L}|0\rangle_{L}&=|101001\rangle, &|1\rangle_{L}|1\rangle_{L}|1\rangle_{L}&=|101010\rangle.
\end{align}
In the case of three-qubit $C_{2}$-$U$ gate, we also use only two neighboring physical qubits to encode a logical qubit and $|a_{3}\rangle=|100011\rangle$ and $|a_{4}\rangle=|101100\rangle$ are as ancillary states.
The Hamiltonian $H_{2}$ for implementing the $C_{2}$-$U$ gate is
\begin{eqnarray}\label{e18}
H_{2}&=&\frac{1}{4}\bigg\{(I_{1}-Z_{1})(I_{4}+Z_{4})
\Big[\Delta_{1}(I_{3}+Z_{3})
+(\Omega_{1}R^{x}_{35}
+\Omega_{2}R^{x}_{36}+\mathrm{H.c.})\Big]\cr\cr
&&+(I_{1}-Z_{1})(I_{3}-Z_{3})
\Big[\Delta_{2}(I_{4}-Z_{4})
+(\Omega_{3}R^{x}_{45}
+\Omega_{4}R^{x}_{46}+\mathrm{H.c.})\Big]\bigg\}\cr\cr
&=&\Big[2\Delta_{1}|a_{3}\rangle\langle a_{3}|+(\Omega_{1}|1\rangle_{L}|1\rangle_{L}|0\rangle_{L}\langle a_{3}|+\Omega_{2}|1\rangle_{L}|1\rangle_{L}|1\rangle_{L}\langle a_{3}|+\mathrm{H.c.})\cr\cr
&&+2\Delta_{2}|a_{4}\rangle\langle a_{4}|+(\Omega_{3}|a_{4}\rangle_{L}\langle 1|_{L}\langle1|_{L}\langle1|+\Omega_{4}|a_{4}\rangle_{L}\langle 1|_{L}\langle1|_{L}\langle0|+\mathrm{H.c.})\Big],
\end{eqnarray}
where the controllable coupling parameters are chosen the same as in the case of $C_{1}$-$U$ gate (see Eq.~({\ref{e10}})). In this way the Hamiltonian in Eq.~({\ref{e18}}) can be rewritten as
\begin{eqnarray}\label{e19}
H_{2}'&=&-2\Omega(\sin\xi|a_{3}\rangle\langle a_{3}|
+\sin\gamma|a_{4}\rangle\langle a_{4}|)\cr\cr
&&+\Omega(\cos\xi|1\rangle_{L}|1\rangle_{L}|+\rangle_{L}\langle a_{3}|
+\cos\gamma|1\rangle_{L}|1\rangle_{L}|-\rangle_{L}\langle a_{4}|+\mathrm{H.c.}).
\end{eqnarray}
The Hamiltonian $H_{2}'$ has the same structure as $H_{1}'$ and the states $|+\rangle_{L}$ and $|-\rangle_{L}$ are the same as that in Eq.~({\ref{e11}}). Similar to the case of $C_{1}$-$U$ gate, it is easy to get the evolution operator associated with $H_{2}$ under the basis $\{|0\rangle_{L}|0\rangle_{L}|0\rangle_{L},$ $|0\rangle_{L}|0\rangle_{L}|1\rangle_{L},$
$|0\rangle_{L}|1\rangle_{L}|0\rangle_{L},$ $|0\rangle_{L}|1\rangle_{L}|1\rangle_{L},$ $|1\rangle_{L}|0\rangle_{L}|0\rangle_{L},$ $|1\rangle_{L}|0\rangle_{L}|1\rangle_{L},$ $|1\rangle_{L}|1\rangle_{L}|0\rangle_{L},$ $|1\rangle_{L}|1\rangle_{L}|1\rangle_{L}\}$
\begin{eqnarray}\label{e20}
\mathcal{U}_{2}(\tau_{2})=\mathrm{Diag}\left[1,1,1,1,1,1,U\right],
\end{eqnarray}
with evolution time satisfying $\Omega\tau_{2}=\pi$. From Eq.~({\ref{e20}}), one can easily find that $\mathcal{U}_{2}(\tau_{2})$ acts as a $C_{2}$-$U$ gate on the states of $S^{L_{2}}$ and $U$ is given by Eq.~({\ref{e05}}). A Toffoli gate, which can perform a NOT operation on the target qubit or not, depending on the states of two control qubits~\cite{ETIJTP8221}, is an important $C_{2}$-$U$ gate. One can get a Toffoli gate by using at least six CNOT gates in principle~\cite{VIQIC099}. Here the Toffoli gate can be directly implemented by utilizing the same parameters in the case of implementing CNOT gate. The decoherence-free and holonomy properties of the gate can now easily be verified. Since the verification exactly parallels the one for the case of $C_{1}$-$U$ gate discussed in the last section and we don't present here.

Now we turn to the implementation of a Fredkin gate, which is another important three-qubit controlled gate that can perform a swap operation on two target qubits or not, depending on the state of the control qubit. In order to achieve the Fredkin gate we consider the following Hamiltonian
\begin{eqnarray}\label{e21}
H_{3}&=&\frac{1}{2\sqrt{2}}\eta(I_{1}-Z_{1})(R^{x}_{35}
-R^{x}_{46}+\mathrm{H.c.})\cr\cr
&=&\eta\frac{1}{\sqrt{2}}(|1\rangle_{L}|1\rangle_{L}|0\rangle_{L}\langle a_{3}|+|1\rangle_{L}|0\rangle_{L}|1\rangle_{L}\langle a_{4}|\cr\cr
&&-|1\rangle_{L}|0\rangle_{L}|1\rangle_{L}\langle a_{3}|-|1\rangle_{L}|1\rangle_{L}|0\rangle_{L}\langle a_{4}|+\mathrm{H.c.})\cr\cr
&=&\eta(|1\rangle_{L}|1\rangle_{L}|0\rangle_{L} -|1\rangle_{L}|0\rangle_{L}|1\rangle_{L})\langle a_{-}|+\mathrm{H.c.},
\end{eqnarray}
where $\eta$ is a controllable coupling parameter and $|a_{-}\rangle=\dfrac{1}{\sqrt{2}}(|a_{3}\rangle-|a_{4}\rangle)$.  Here the encoding is as the same as the situation in $C_{2}$-$U$ gate (see Eq.~({\ref{e18}})). The Hamiltonian $H_{3}$ is in the $\Lambda$-type with
ancillary state $|a_{-}\rangle$ at the top while the logical qubit states $|1\rangle_{L}|1\rangle_{L}|0\rangle_{L}$ and $|1\rangle_{L}|0\rangle_{L}|1\rangle_{L}$ at the bottom. The state orthogonal to $|a_{-}\rangle$ is denoted as $|a_{+}\rangle=\dfrac{1}{\sqrt{2}}(|a_{3}\rangle+|a_{4}\rangle)$ and it decouples from the evolution of the system. The subspace spanned by $\{|a_{+}\rangle$, $|a_{-}\rangle\}$ is the same to that by $\{|a_{3}\rangle$, $|a_{4}\rangle\}$. When the evolution time $\tau_{3}$ meets
$\eta\tau_{3}=\pi/\sqrt{2}$, the resulting evolution operator in the basis
$\{|0\rangle_{L}|0\rangle_{L}|0\rangle_{L},$ $|0\rangle_{L}|0\rangle_{L}|1\rangle_{L},$
$|0\rangle_{L}|1\rangle_{L}|0\rangle_{L},$ $|0\rangle_{L}|1\rangle_{L}|1\rangle_{L},$ $|1\rangle_{L}|0\rangle_{L}|0\rangle_{L},$ $|1\rangle_{L}|0\rangle_{L}|1\rangle_{L},$ $|1\rangle_{L}|1\rangle_{L}|0\rangle_{L},$ $|1\rangle_{L}|1\rangle_{L}|1\rangle_{L}\}$ is given by
\begin{eqnarray}\label{e22}
\mathcal{U}_{3}(\tau_{3})=\mathrm{Diag}\left[1,1,1,1,
\left(\begin{array}{cccc}
1~&~0~&~0~&~0\\
0~&~0~&~1~&~0\\
0~&~1~&~0~&~0\\
0~&~0~&~0~&~1\end{array}\right)\right].
\end{eqnarray}
One can find from Eq.~({\ref{e22}}) that $\mathcal{U}_{3}(\tau_{3})$ acts as a Fredkin gate on the states in the logic subspac $S^{L_{2}}$ and its decoherence-free and holonomy properties can be demonstrated easily. In this way we implement a non-adiabatic holonomic three-qubit Fredkin gate in DFS with three-body interaction.

\section{Discussion and Conclusions}\label{sec3}
So far, we have succeeded in constructing $C_{1}$-$U$, $C_{2}$-$U$, and Fredkin gates. We now introduce a few concepts from differential geometry to understand the nature of the above holonomic gates. The set of $K$-dimensional subspaces of an $N$-dimensional Hilbert space is a Grassmann manifold $G(N; K)$. The closed path $\mathcal{C}$ of $K$-dimensional subspaces is a loop in $G(N; K)$. We now consider the holonomic gates described above. The $C_{1}$-$U$, $C_{2}$-$U$, and Fredkin gates are associated with loops in $G(4; 2)$~\cite{VCENJP1416}, where the Hilbert spaces relevant for the holonomy is spanned by $\{|a_{1}\rangle, |a_{2}\rangle, |1\rangle_{L}|0\rangle_{L}, |1\rangle_{L}|1\rangle_{L}\}$, $\{|a_{3}\rangle, |a_{4}\rangle, |1\rangle_{L}|1\rangle_{L}|0\rangle_{L}, |1\rangle_{L}|1\rangle_{L}|1\rangle_{L}\}$, and $\{|a_{3}\rangle, |a_{4}\rangle, |1\rangle_{L}|0\rangle_{L}|1\rangle_{L}, |1\rangle_{L}|1\rangle_{L}|0\rangle_{L}\}$, respectively. However, the previous schemes were almost associated with loops in $G(3; 2)$~\cite{EDLBMKNJP1214}. It is worth noting that the schemes proposed here can be generalized. For the $C_{1}$-$U$ gate between the $m$th and the $n$th logic qubits, the Hamiltonian has the same structure as $H_{1}$ but with the exchanging $R^{x}_{13}\rightarrow R^{x}_{2m-1,2n-1}$, $R^{x}_{14}\rightarrow R^{x}_{2m-1,2n}$, $R^{x}_{23}\rightarrow R^{x}_{2m,2n-1}$, $R^{x}_{24}\rightarrow R^{x}_{2m,2n}$,  $(I_{1}+Z_{1})\rightarrow (I_{2m-1}+Z_{2m-1})$, and $(I_{2}+Z_{2})\rightarrow (I_{2m}+Z_{2m})$. For the $C_{2}$-$U$ gate between the $m$th, $n$th, and $l$th logic qubits, the Hamiltonian has the same structure as $H_{2}$ but with the exchanging $R^{x}_{35}\rightarrow R^{x}_{2n-1,2l-1}$, $R^{x}_{36}\rightarrow R^{x}_{2n-1,2l}$, $R^{x}_{45}\rightarrow R^{x}_{2n,2l-1}$, $R^{x}_{46}\rightarrow R^{x}_{2n,2l}$, $(I_{1}-Z_{1})\rightarrow (I_{2m-1}-Z_{2m-1})$, $(I_{3}+Z_{3})\rightarrow (I_{2n-1}+Z_{2n-1})$, and $(I_{4}+Z_{4})\rightarrow (I_{2n}+Z_{2n})$. At last, for the Fredkin gate between the $m$th, $n$th, and $l$th logic qubits, the Hamiltonian has the same structure as $H_{3}$ but with the exchanging $R^{x}_{35}\rightarrow R^{x}_{2n-1,2l-1}$, $R^{x}_{46}\rightarrow R^{x}_{2n,2l}$ and $(I_{1}-Z_{1})\rightarrow (I_{2m-1}-Z_{2m-1})$.

In conclusion, we have proposed schemes for implementing $C^{1}$-{\it U}, $C^{2}$-{\it U}, and Fredkin gates directly by using non-adiabatic holonomy in DFSs. Our schemes combine the coherence stabilization virtues of DFSs and the built-in fault tolerance of holonomic control. These gate operations can be implemented in a high speed which avoids the extra errors and decoherence involved in adiabatic case due to long time evolution. Moreover, the resource cost for the DFSs encoding is minimal by using only two neighboring physical qubits undergoing collective dephasing to encode a logical qubit.

\begin{center}
{\small {\bf ACKNOWLEDGMENTS}}
\end{center}

This work was supported by the National Natural Science Foundation of China under
Grant Nos. 11264042, 11465020, 61465013, 11165015, and 11564041.

\end{document}